# Adiabatic embedment of nanomechanical resonators in photonic microring cavities

Chi Xiong[1], Wolfram Pernice[1], Mo Li[1], Michael Rooks[2] and Hong X. Tang[1]

[1] *Department of Electrical Engineering, Yale University*

*New Haven, CT 06520, USA*

[2] *Yale Institute of Nanoscience and Quantum Engineering,*

*15 Prospect St, New Haven, CT 06520 USA*

(May 15, 2010)

**Abstract:**

We report a circuit cavity optomechanical system in which a nanomechanical resonator is adiabatically embedded inside an optical ring resonator with ultralow transition loss. The nanomechanical device forms part of the top layer of a horizontal silicon slot ring resonator, which enables dispersive coupling to the dielectric substrate via a tapered nanogap. Our measurements show nearly uncompromised optical quality factors ($Q$) after the release of the mechanical beam.

a) Electronic mail: hong.tang@yale.edu

Nanomechanical resonators, size-matched to nanophotonic components can be efficiently actuated by gradient optical forces[1,2,3] and hence enable nanophotonic device functionalities ranging from optomechanical cooling to ultra-sensitive all-photonic detection. To enhance the detection sensitivity, high finesse cavities[4,5] can be employed. In such a system both the detection sensitivity and the optical force are expected to be enhanced by a factor related to the quality factor ($Q$) of the optical cavity. A high optical $Q$ factor is also greatly desirable for achieving optomechanical cooling and amplification in a photonic circuit. However, when a nanomechanical resonator is inserted into a photonic circuit, additional optical loss often occurs at the clamping points of the waveguides and compromises the optical $Q$. To overcome this loss, several approaches have been developed, including the use of multimode interference couplers[1] and photonic crystal waveguides[6] to match the modes in the free-standing resonators to the modes in the substrate supported waveguide. While such designs lead to improved mechanical clamping, the resulting finite optical insertion loss still prevents the formation of a high $Q$ circuit cavity with embedded nanoelectromechanical systems (NEMS).

The fabrication of nanomechanical resonators typically involves removing an underlying sacrificial layer to form free-standing waveguides, resulting in a finite substrate-waveguide separation. For instance, for a waveguide beam to be released from silicon on insulator (SOI) substrates by wet etching, the minimum obtainable gap is about 300 nm[1]. Precise control of the final etch depth by tuning the etching time is difficult due to the isotropic nature of the etch process. Near the clamping points of the beam, the resulting step of several hundred nanometers in the dielectric substrate causes the effective mode index of the waveguide to change abruptly and thus considerable insertion loss occurs as a consequence of the non-adiabatic mode propagation.

In this work, we employ horizontal slot waveguides[7,8] and ring resonators to circumvent clamping losses and realize optomechanical resonators without compromising the optical $Q$. The slot waveguides are fabricated from an α-Si/SiO$_2$/Si layer structure, comprising an amorphous silicon (α-Si) top layer and a thin SiO$_2$ slot layer on top of an SOI substrate, as illustrated in the inset of Figure 1 (a). We release a nanomechanical beam from the amorphous silicon layer by wet etching the thin SiO$_2$ slot layer. By comparing the optical quality factors before and after the releasing process, we find that in our horizontal slot waveguide system the optical insertion loss occurring at the boundary of the nanomechanical beam is negligible.

Our fabrication process is realized on SOI substrates with a 110 nm thick silicon layer and a 3 μm buried oxide (BOX) layer. 80 nm silicon dioxide and 200 nm amorphous silicon were deposited using plasma enhanced chemical vapor deposition (PECVD). Nanophotonic waveguides of 650 nm width as well as input/output grating couplers were patterned by electron beam lithography and etched into the amorphous silicon layer in an inductively coupled chlorine plasma (ICP) etcher. The etch windows for the wet release of the nanomechanical resonator were subsequently defined by photolithography and the beams were released from the substrate by wet etching using buffered oxide etchant (BOE). Figure 1 (a) shows the optical microscope image of a fabricated device and Figure 1 (b) shows a zoomed-in view of the released beam as observed by scanning electron microscope (SEM).

The 10 μm long beam buckles up from the substrate as a result of high compressive stress, which was introduced during the PECVD process. Under normal circumstances, the wet etching process causes stiction of NEMS beams to the substrate when the released gap is small. Here the up-buckled state is mechanically stable and therefore avoids the stiction problem. The buckling profile is imaged by atomic force microscope (AFM) as shown in Figure 2 (a), from which buckling of 450 nm is measured. Using an analytical model[9] for stressed doubly clamped beams, we can further determine the compressive stress to be about 1 GPa, consistent with values measured on bare wafers. Finite element method (FEM) simulation of the buckled beam shows a fundamental out-of-plane mechanical resonance at 19.2 MHz. The inset in Figure 2 (c) displays the simulated mode shape and the strain distribution. Additionally, we show in Figure 2 (b) that the released waveguide supports continuous optical modes. Using finite-difference time-domain (FDTD) simulations we estimated the insertion loss of the buckled beam structure in dependence of buckling amplitude as shown in Fig. 2 (c). When the buckling amplitude is small, the FDTD results suffer from computational uncertainties, because the numerical engine is unable to distinguish between loss values differing by less than 0.05 dB. Therefore we scan a wide range of buckling amplitudes and extrapolate from the results at wider gaps to the actual buckling amplitude. At a buckling amplitude of 450 nm, the insertion loss is projected to be around 0.1±0.05 dB for 650 nm wide waveguides.

To verify the numerical predictions we compared the measured optical transmission spectrum before and after the release of the beam, as shown in Figure 1 (c). We fitted ten resonances from 1540 nm to 1565 nm to Lorentzian lineshapes. Before release, we found an average quality factor of 19800±1060. After the release process the average $Q$ is 18800±990. Therefore we estimate that the insertion loss induced by the release of the beam to be 0.04 dB±0.02 dB (1%±0.5%). By launching a probe light with a frequency slightly detuned from cavity resonances, we are able to observe the thermomechanical motion of the beam in vacuum at 1 mTorr. Figure 3 (a) displays the measured noise spectrum in the probe signal, showing mechanical resonance frequency at 19.97 MHz which agrees well with the fundamental out-of-plane mode frequency of the device as expected from FEM simulation. Fitting the resonance yields a mechanical quality factor of ~ 870, slighter lower than the value obtained in an unbuckled, doubly clamped beam[6]. The displacement sensitivity is 11 fm/√Hz, achieved with moderate probe laser power of 1 mW in the input waveguide.

To further characterize the nanomechanical response of the suspended amorphous silicon beam driven by the gradient optical force, we use the pump and probe scheme presented in our earlier work. In addition to the detuned probe light, we apply an intensity modulated pump light with 5 mW input power. Taking into account the insertion loss of the input grating coupler we estimate the optical power in the input waveguide to be 1 mW. On resonance, the optical power on the nanomechanical resonator will be enhanced by the finesse of the ring, which is found to be approximately 33 in our device. Figure 3(b) shows the amplitude and phase response of the beam. The resonance response of the beam is measured as a function of the laser modulation amplitude showing a linear relationship.

The optical force will increase rapidly with decreasing coupling gap between the beam and the underlying dielectric substrate[10]. By further device optimization and stress engineering, it is possible to prevent the beams from buckling up[11] and hence the coupling gap between the beam and the substrate will only be determined by the sacrificial layer thickness, which can be predefined to be as thin as 10 nm. Thus the greatly enhanced light matter interaction in the small air gaps makes the horizontal slot waveguide an attractive candidate for ultra sensitive readout of high frequency nanomechanical properties.

In conclusion, we demonstrated very low loss insertion of a nanomechanical beam into a ring cavity, confirmed by FDTD simulations and optical transmission measurements. Our results show that disturbance of the optical mode due to the nanomechanical beam can be minimized by optimizing mechanical and optical design. We anticipate that the combination of horizontal slot waveguides with nanomechanical beams will find applications in ultrasensitive detection and cavity optomechanical cooling and amplification of nanomechanical devices.

This work was supported by a seedling program from DARPA/MTO and the DARPA/MTO ORCHID program through a grant from AFOSR. W.H.P. Pernice would like to thank the Alexander-von-Humboldt foundation for providing a postdoctoral fellowship. H.X.T acknowledges support from a Packard Fellowship in Science and Engineering and a CAREER award from the National Science Foundation. We thank Yale School of Engineering and Applied Sciences cleanroom managed by Michael Power. The project makes use of electron beam lithography facilities at Brookhaven national laboratory, which is supported by the U.S. Department of Energy, Office of Basic Energy Sciences, under Contract No. DE-AC02-98CH10886.

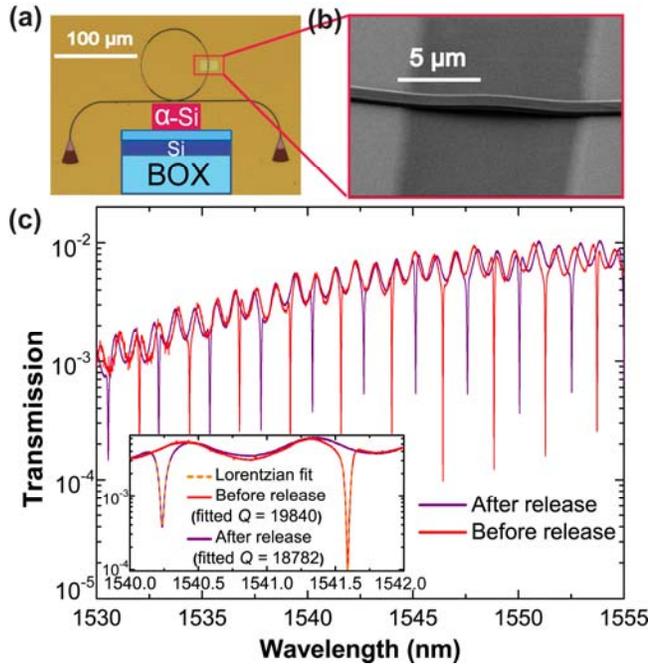

Figure 1.

(a) Optical microscope image of the 40 μm radius ring resonator with inset showing the cross-sectional schematic of the horizontal slot waveguide structure with 80 nm $SiO_2$ slot.

(b) The 10 μm long amorphous silicon beam released from the substrate as observed in the SEM. The buckling is a result of residual compressive stress.

(c) Optical transmission spectra before and after the releasing process. Inset shows the fitted $Q$ factors for a cavity resonance around 1540 nm wavelength, before and after the release.

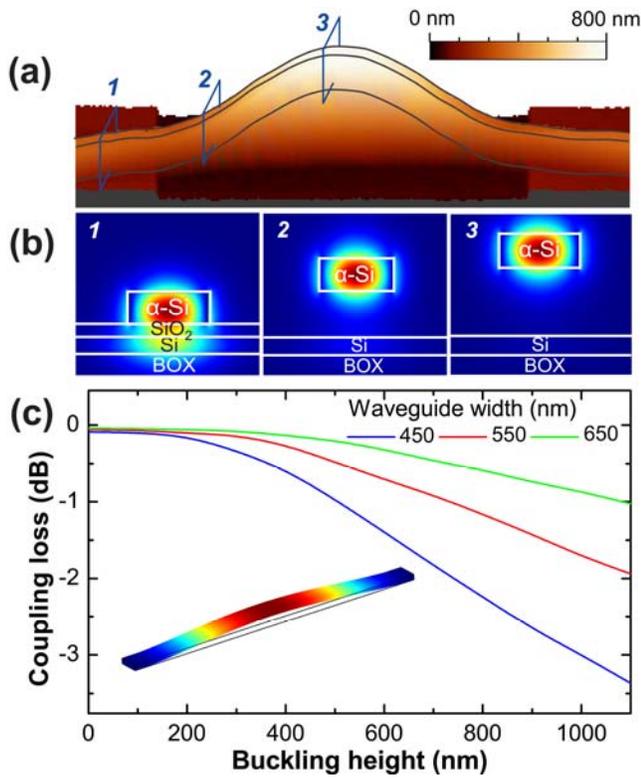

Figure 2.

(a) A high resolution AFM image of the released beam showing buckling of 450 nm with the contour of the nanomechanical beam outlined;

(b) Simulated fundamental optical mode profile along three cut locations as labeled in panel (a).

(c) FDTD simulations of the propagation loss along the buckled beam versus the buckling amplitude as a function of the waveguide width. The inset shows the FEM simulation of the fundamental out-of-the-plane mechanical mode shape of the beam with the stress distribution displayed in color.

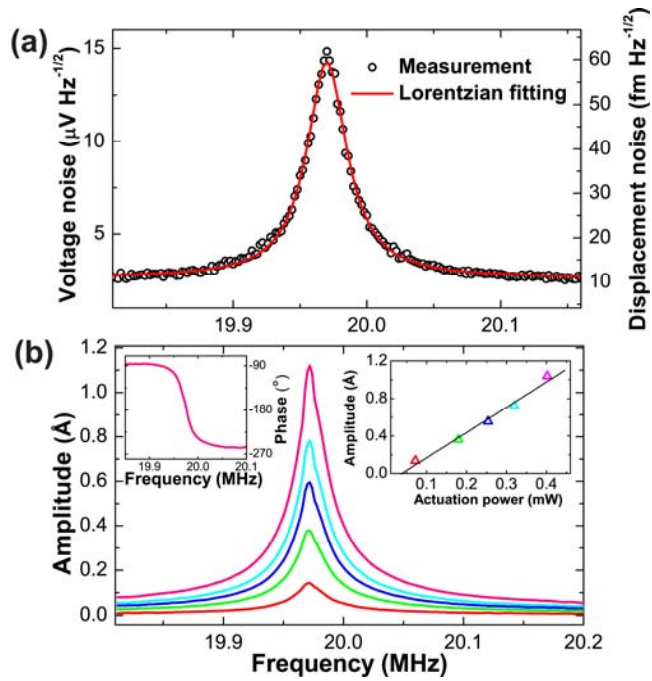

Figure 3.

(a) Thermomechanical noise spectrum density and calibrated displacement sensitivity. Measurement was taken in a vacuum of 1 mTorr;

(b) The frequency response of the beam resonances at various driving amplitude. The left inset shows the phase response at 0.4 mW actuation power.

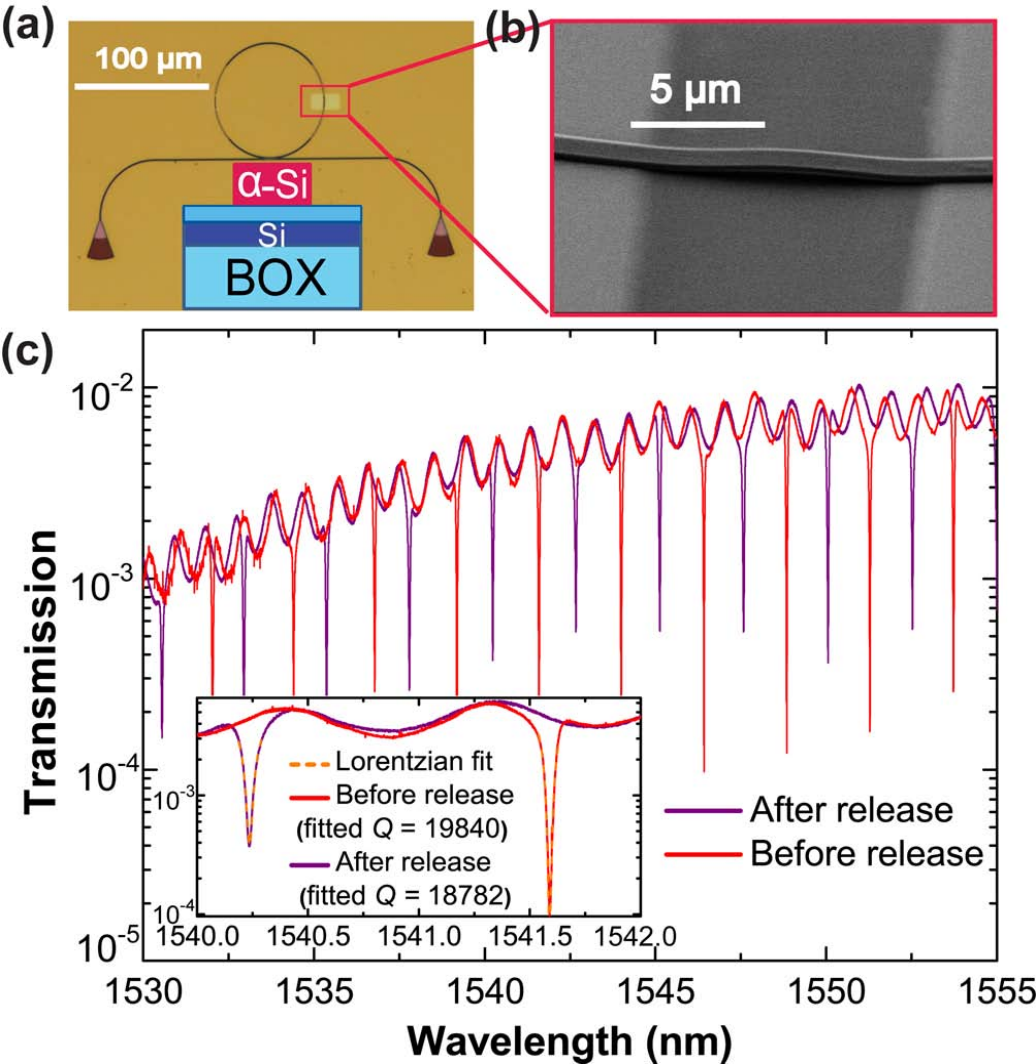

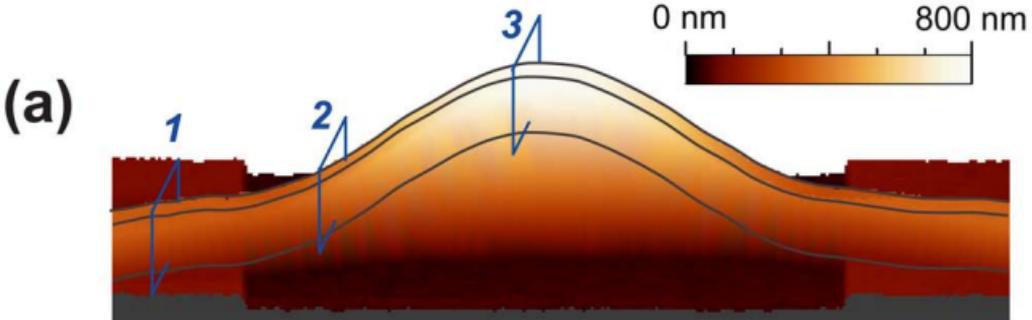
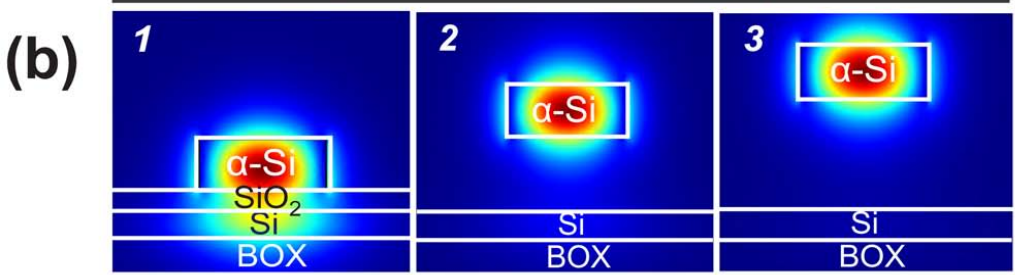
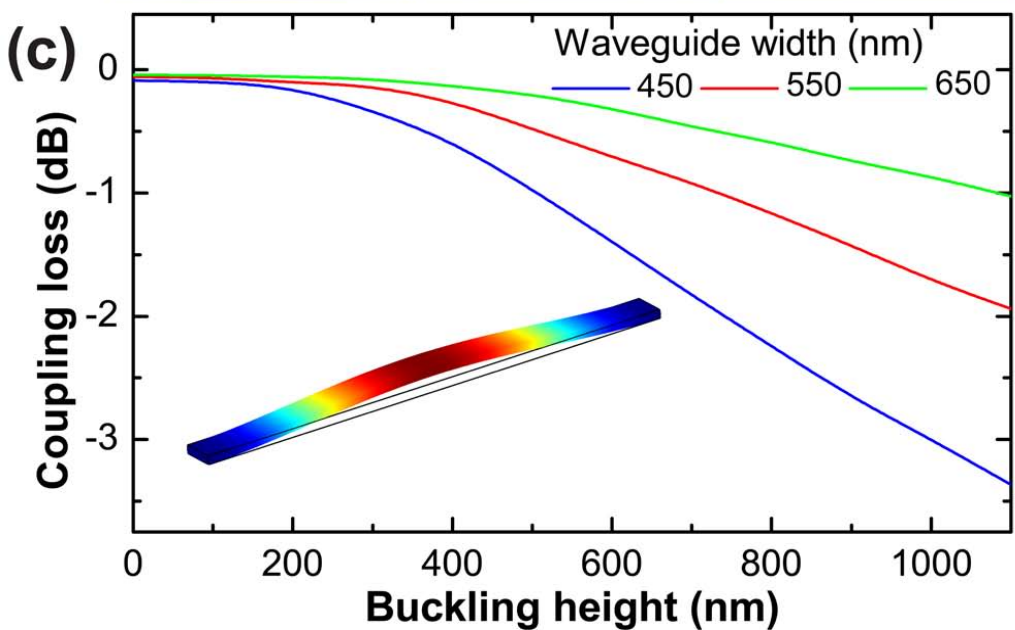

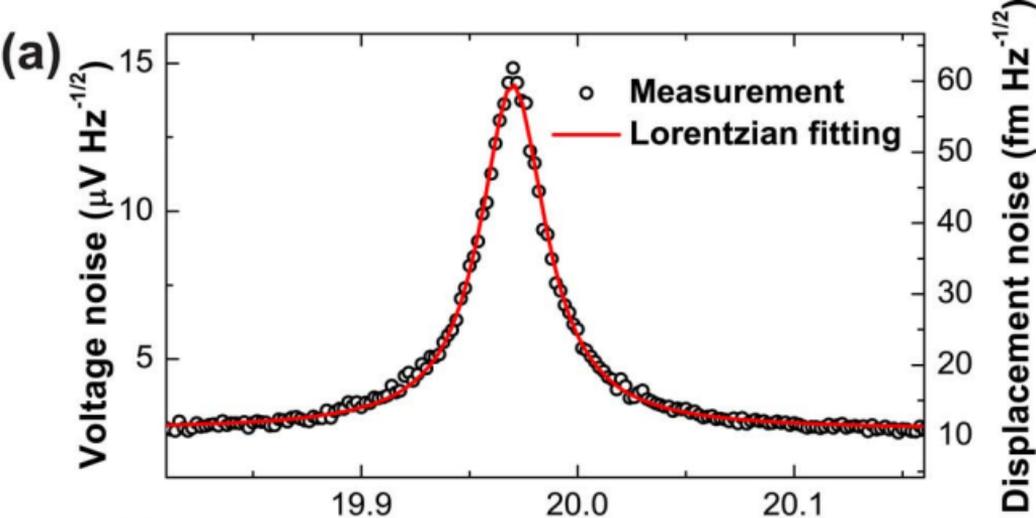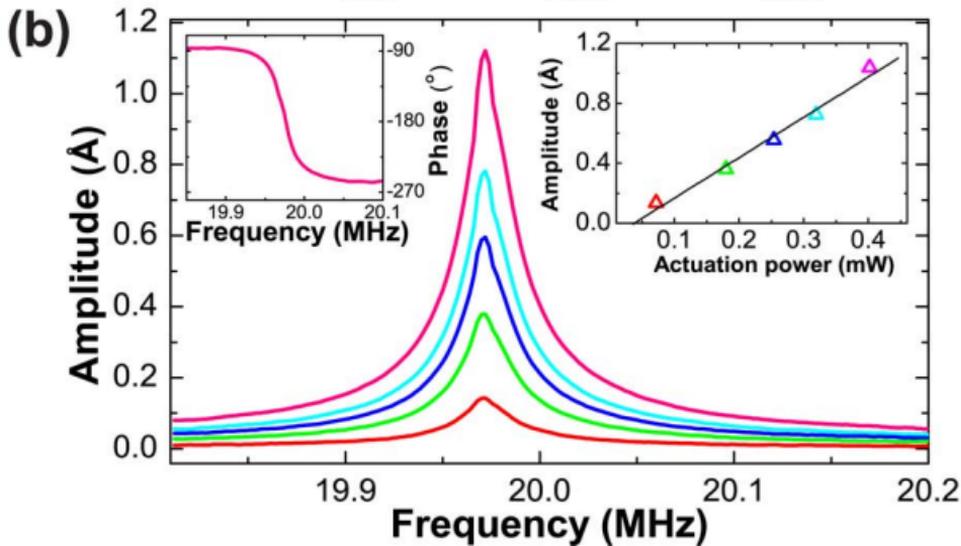